\documentclass[twocolumn,english,aps,prb]{revtex4}
\usepackage{amsmath}
\usepackage{graphicx}

\def\beq{\begin{equation}}
\def\eeq{\end{equation}}
\def\beqa{\begin{eqnarray}}
\def\eeqa{\end{eqnarray}}

\def\e{\epsilon}

\def\e{\epsilon}
\def\cH{{\mathcal H}}

\def\Tr{\mathrm{Tr}}

\def\si{\sigma}

\def\o{\omega}

\def\nonum{\nonumber \\}

\def\nonum{ \nonumber \\}

\begin{document}  
\author{Yonatan Dubi}
\email{jdubi@bgu.ac.il}
\affiliation{Department of Chemistry and the Ilze-Katz Institute for Nano-Scale Science and Technology, Ben-Gurion University of the Negev, Beer-Sheva 84105, Israel}
\title{Dynamical coupling and negative differential resistance from interactions across the molecule-electrode interface in molecular junctions }
\date{\today}
\begin{abstract}
Negative differential resistance - a decrease in current with increasing bias voltage - is a counter-intuitive effect that is observed in various molecular junctions. 
Here, we present a novel mechanism that may be responsible for such an effect, based on strong Coulomb interaction between electrons in the molecule and electrons on the 
atoms closest to the molecule. The Coulomb interaction induces electron-hole binding across the molecule-electrode interface, resulting in a renormalized and enhanced 
molecule-electrode coupling. Using a self-consistent non-equilibrium Green's function approach, we show that the effective coupling is non-monotonic in bias
voltage, leading to negative differential resistance. The model is in accord with recent experimental observations that showed a correlation between the 
negative differential resistance and the coupling strength. We provide detailed suggestions for experimental tests which may help to shed light on the origin of the negative differential 
resistance. Finally, we demonstrate that the interface Coulomb interaction affects not only the I-V curves but also the thermoelectric properties of molecular junctions.                                                                                                                                          
\end{abstract}
\maketitle   


\section{Introduction}
Single-molecule junctions offer a rich variety of effects, including switching, rectification, photo-activity and thermoelectricity
\cite{Tao2006,Tsutsui2012,Song2011,Selzer2006}. Especially intriguing - due to its counter-intuitive nature - is the negative differential resistance (NDR) effect: When a 
molecular junction is placed under a voltage bias, the current is typically a monotonically increasing function of bias (resulting in the usual positive differential 
resistance). However, in certain situations, the current decreases with increasing bias, resulting in the so-called NDR. This effect has been observed in a large variety of 
metal-molecule-metal junctions, including organic and metallo-organic molecules and even DNA junctions \cite{Zhou2013,Le2003,Khondaker2004,Kang2010,Chen2000,Chen1999,Migliore2011,Burkle2012}. The 
origin of the NDR, being unclear, has spurred a large number of theoretical studies. The main  
sources for NDR discussed in the literature are voltage-dependent electron-phonon interactions \cite{Galperin2005,Han2010,Hartle2011} and a potential-drop-induced shifting of the molecular orbitals \cite{Wang2012,Long2008,Zimbovskaya2008,Lu2005,Garcia-Suarez2012,Wang2006,Dalgleish2006,Thygesen2008}. In addition, recent studies indicate that NDR may originate from image-charge effects \cite{Hedegard2005,Kaasbjerg2011} and a change in the effective coupling  between the molecular levels and the electrodes \cite{Huang2012,Pati2008}. 

In a recent experiment \cite{Zhou2013}, the transport properties of a molecular junction composed of a symmetric metallo-oranic molecule between two gold electrodes were 
studied in parallel to force measurements on the molecular junction, which determine the effective coupling strength between the molecule and the electrodes \cite{Zhou2011}.  
The authors found that NDR appears only for junctions with intermediate coupling, at a relatively low bias of $\sim 0.64$ eV, which is much smaller than the HOMO-LUMO gap; yet 
some features in the local density of states (LDOS) seem to appear at these energies. A major finding described in Ref.~\cite{Zhou2013} was that the peak current is correlated with a maximum in the molecule-electrode coupling.

Here we suggest a novel mechanism for the NDR, which is in line with the above experimental findings. The mechanism that we suggest relies on the assumption that the Coulomb 
interactions between electrons on the molecule and electrons on the electrode (residing on the atoms connected to the molecule) are significant and are not 
screened out by the free electrons in the electrode. Theoretical simulations indicate that when molecular junctions are formed, atoms from the electrode may extend towards 
the molecule \cite{Burkle2012a,Demir2011,Kim2013}, thereby reducing the screening in the bridging atoms. The effect of such interface Coulomb interactions has recently been studied 
theoretically, showing both that the 
magnitude of these interactions can be large (a few electron volts) and that they strongly affect the molecular orbitals \cite{Strange2012}, via the formation of image charges. The 
importance of considering image charges in calculating I-V characteristics was pointed out in Ref. ~\cite{Darancet2012}. 

The model we consider and the mechanism we propose are depicted in Fig.~\ref{fig1}. The system under consideration is composed of a single molecular level, say the lowest unoccupied molecular level (LUMO), coupled
to the electronic states in the electrode. We consider Coulomb interactions both on the molecular level (leading to the well-known Coulomb blockade \cite{DiVentra2008}) and 
between the molecular level and the electrons on the atom connected to the molecular bridge (bridging atom). When an electron hops from the bridging atom to the LUMO, a hole is left behind. 
Due to the Coulomb interactions, the electron on the LUMO and hole are correlated, and this correlation renormalizes the hopping matrix element between the LUMO and the bridging atom, 
thereby increasing the coupling between the molecule and the electrode (as we explicitly show in what follows). In the presence of a bias voltage, the rate of electron hopping between the electrode and the LUMO changes, which induces a 
bias-dependent normalization of the coupling. When the bias reaches the LUMO energy level, the current is maximal and thus the renormalization is maximal, and as the resonance is passed, the 
effective coupling is reduced and (self-consistently) the current is reduced, resulting in NDR.

\section{Model and effective coupling} 
\begin{figure}[h!]\centering
\includegraphics[width=7truecm]{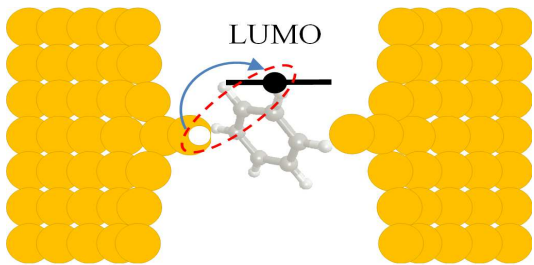}
\caption{Schematic representation of the molecular junction under consideration. When an electron hops from the bridging site to the LUMO, a hole is left behind. Due to the 
interface Coulomb interaction, the electron on the LUMO and the hole on the bridging site are attracted, forming a bound exciton. This process renormalizes and enhances the 
tunneling matrix element between the bridging site and the LUMO, thereby affecting the current through the junction. }
\label{fig1}
\end{figure}

The starting point is the typical Hamiltonian for a molecular junction \cite{Tsutsui2012,Selzer2006,DiVentra2008}:
\beqa
\cH &=& \cH_L+\cH_R+\cH_M+\cH_T+\cH_{I} \nonum
\cH_{L,R} &=& \sum_{k\si} \e_k c^{\dagger}_{k\si,L/R} c_{k\si,L/R} \nonum 
\cH_M &=& \e_0  \sum_\si d^{\dagger}_\si d_\si + U_0 d^{\dagger}_\uparrow d_\uparrow d^{\dagger}_\downarrow d_\downarrow  \nonum 
\cH_T &=& -t\sum_\si \left( c^{\dagger}_{\si,L} d_\si+ c^{\dagger}_{\si,R} d_\si+\textrm{h.c.} \right) \nonum \cH_I &=&  U_1 \left( \hat{n}_L \hat{n}_d+\hat{n}_R \hat{n}_d \right)
\label{Hamiltonian} \eeqa 
where $c^{\dagger}_{k,L/R}$ creates an electron with momentum $k$ and spin $\si$ at the left/right electrode, and $d^\dagger_\si$ creates an electron with spin $\si$ 
at the LUMO. The first three terms are a typical description of transport through molecular junctions, with metallic electrodes, molecular orbital energy $\e_0$, and tunneling matrix element $-t$ 
connecting the molecule and the electrodes. The tunneling from 
the electrodes to the molecule occurs through the left and right bridging atoms, defined by the electron creation operators
 $c^{\dagger}_{\si,L} ,c^{\dagger}_{\si,R} $ for the 
left and right electrodes, as depicted in Fig.~\ref{fig1}. The interaction term with $U_0$ describes the molecular charging energy and leads to the Coulomb blockade. 

The last term in the Hamiltonian describes the Coulomb interaction between the electron on the molecule and an electron in the 
bridging site, where $U_1$ the interaction strength and $\hat{n}_L =\sum_\si c^{\dagger}_{\si,L}c_{\si,L}$ and $\hat{n}_d =\sum_\si d^{\dagger}_{\si}d_{\si}$ are the 
electron number operators on the molecule and the bridging site, respectively. In Ref.~\cite{Thygesen2009}, such an interaction term was considered, and its effect on the molecular 
level renormalization was studied using various theoretical approaches theoretical in equilibrium conditions.  

To study the effect of the interface Coulomb interaction described in $\cH_I$, we introduce a mean-field decoupling for the interaction term 
\beq
c^{\dagger}_{L,\si}c_{L,\si} d^{\dagger}_{\si}d_{\si}\approx -\langle c^\dagger_{L,\si} d_\si \rangle d^\dagger_\si c_{L,\si}-\langle  d^\dagger_\si c_{L,\si} \rangle c^\dagger_{L,\si} d_\si ~~\label{MeanField}
\eeq
where for the sake of clarity, we discuss the interaction with the left electrode and later generalize the expressions to include the right electrode.This decoupling scheme 
renormalizes the tunneling Hamiltonian into \beq \cH_T=-\tilde{t}\sum_\si \left( c^{\dagger}_{\si,L} d_\si+ c^{\dagger}_{\si,R} d_\si+\textrm{h.c.} \right)~,\eeq introducing 
effective tunneling matrix elements $\tilde{t}_{L,d}=t+\tau_{L,d},~\tilde{t}_{d,L}=t+\tau_{d,L}$
 where $\tau_{L,d}=U_1 \langle c^\dagger_{L,\si} d_\si \rangle,~ \tau_{d,L}=U_1 \langle  d^\dagger_\si c_{L,\si} \rangle$. 

Before proceeding, several remarks are in order. First, note that we use the term "renormalization" to describe an interaction-induced change in the coupling, even if the interaction is treated on the mean-field level. Second, we point out that the 
interaction term can also be decoupled diagonally, resulting in a shift of the local energy levels in the molecule due to the density in the bridging site and vice versa. While this term may be (and probably is) finite 
and important, we choose here to ignore it so as to isolate the effect of the tunneling renormalization and its bias dependence. Such a separation between the different mechanisms 
cannot be introduced cannot be introduced in, e.g. first principle calculations \cite{Pati2008}, where only the wave functions (or more accurately the Kohn-Sham orbitals) and the total current are calculated, and the mechanism for NDR needs to be inferred from them.

 Third, we note that while a decoupling of $\cH_I$ to spin-flip 
processes is formally possible, it is evident that the self-consistent solution for such processes results in vanishing terms, i.e., the Coulomb interaction cannot induce spin-flip-tunneling process. Finally, we point out the similarity between the decoupling scheme of Eq.~\ref{MeanField} and that of exciton interactions in bilayer systems \cite{Dubi2010}. Indeed, the process we describe above can be 
envisioned as a formation of a bound exciton at the molecule-electrode interface, with $U_1$ being the exciton binding energy. In Ref.~\cite{Dubi2010} the interplay between exciton tunneling and binding was 
demonstrated, in similarity with the results presented in this study.

To evaluate the tunneling matrix elements, we use the non-equilibrium Green's function formalism and write $ \langle c^\dagger_{L,\si} d_\si \rangle=-i \int \frac{d\omega}{2 \pi} G^<_{Ld,\si}(\omega),~ \langle  d^\dagger_\si c_{L,\si} \rangle =-i \int \frac{d\omega}{2 \pi} G^<_{dL,\si}(\omega)$, where $G^<$ are the equal 
times lesser Green's functions \cite{DiVentra2008}. The Green's function may be evaluated via the Dyson equations for the Keldysh functions \cite{Meir1992}, and the full details of the 
derivation are provided in the appendix. The final result of the calculation is a set of self-consistent equations for the tunneling matrix elements, the effective level broadening, and the 
Green's function of the electrons in the molecule, 

\begin{subequations}
\begin{align}
   g^r_\si(\omega) &= \frac{\omega-\e_0-U_0 (1-n_{\bar{\si}})}{(\omega-\e_0)(\omega-\e_0-U_0)} \label{subeq1}\\
   G^r_\si(\omega) &= \left(  g^r_\si(\omega)^{-1}- i \Gamma  \right)^{-1} \label{subeq2}\\
    \Gamma &= \rho_0 \tilde{t}^2 ~,~\tilde{t}=t+\tau \label{subeq3}\\
    \tau &= \frac{1}{2} U_1 \tilde{t} \rho_0 \int_W d\omega (f_L(\omega)+f_R(\omega)) \sum_\si\Re G^r_\si (\omega) ~~. \label{subeq4}  
\end{align}
\end{subequations}

Here $g^r_\si(\omega)$ is the Green's function in the molecule without the electrodes, which is obtained from the equations of motion method \cite{Souza2007,Meir1991} and captures the Coulomb blockade. $G^r_\si$ is the retarded Green's function of electrons in the molecule, $\rho_0$ is the DOS at the bridging site, $f_{L,R}=\frac{1}{1+\exp \left( \frac{\omega-\mu_{L,R}}{T_{L,R}}\right) }$ are the 
Fermi distributions in the left/right electrodes (with the corresponding chemical potentials and temperatures, which may be different), $W$ is the electrode band width, and $\Gamma$ is the level broadening due to the electrodes (taken in the wide-band limit \cite{DiVentra2008}), renormalized by the affective tunneling matrix elements $\tilde{t}=t+\tau$. In what follows the bare chemical potentials are set as the zero of energy, and the right chemical potential is shifted be the bias voltage $V$. The self-consistent procedure is then the following: starting from an initial value for the parameters $\Gamma$ and $\tau$, the Green's function $g^r_\si(\omega)$ and $G^r_\si$  are calculated using Eq.~(\ref{subeq1}-\ref{subeq2}). From the Green's function, $\tau$ and $\Gamma$ are calculated using Eq.~(\ref{subeq3}-\ref{subeq4}). Once they are obtained, the Green's functions are re-calculated with the new value of $\Gamma$. The process is then repeated until self-consistency is achieved, i.e. until the changes in the Green's functions and $\Gamma$ between successive iterations are smaller than some numerical tolerance.

From Eq.~(\ref{subeq1}-\ref{subeq4}), one can see that even at zero bias the tunneling 
element is renormalized, since even at zero bias there is hybridization between the LUMO and the electrode states. In addition, note that in the absence of tunneling (i.e., $t=0$) the self-consistent 
$\tau$ always vanishes, i.e., the Coulomb interaction cannot induce tunneling but can only enhance it (this is the reason for the vanishing of the spin-flip terms). 
Once the the Green's functions are obtained, the current can be calculated via \cite{Meir1992,DiVentra2008} (assuming symmetric electrodes) 
\beq
I=\frac{e}{h}\int_W d\omega (f_L(\omega )-f_R(\o )) \Tr[\Gamma G^r_\si(\omega ) \Gamma G^a_\si(\omega )]~~.\label{MeirWingreen}
\eeq

\section{Results} 
For the numerical calculation, we chose numerical values for the different parameters that are representative of molecular junctions, but we stress that we are aiming at 
a qualitative demonstration of the NDR rather than a quantitative fit to experiments. We used $\e_0=0.55 $ eV for the LUMO, $t\approx 0.3$ eV for the tunneling element, $\rho_0=0.1~\textrm{eV}
^{-1}$ for the DOS (resulting in level broadening of $\Gamma\sim 10$ meV), $W=8$ for the bandwidth and $T=300$ K for the temperature. The chemical potentials were set as the zero energy, and 
the right electrode Fermi distribution was shifted by the bias voltage $V$ or a temperature difference $\Delta T$. The Coulomb energies were extracted from an Ohno parametrization of the 
electrostatic interaction
 \cite{Hedegard2005}, where the Coulomb interaction term depends on the distance between two electrons as $U(r)=14.4/\sqrt{(14.4/U_0)^2+r^2}$ (with $r$ in Angstroms). A 
typical value $U_0=11.26$ eV and a distance of $5$ \AA ~ from the molecule to the bridging site yield a typical $U_1=2.6$ eV which, if not screened, is a rather strong 
interaction. We point that the onsite Coulomb interaction term $U_0$ is not especially important for the mechanism of NDR we suggest here, yet we keep $U_0$ finite to 
agree with parameters presented in previous literature.\cite{Hedegard2005}

\begin{figure}[h!]
\centering
\includegraphics[width=7truecm]{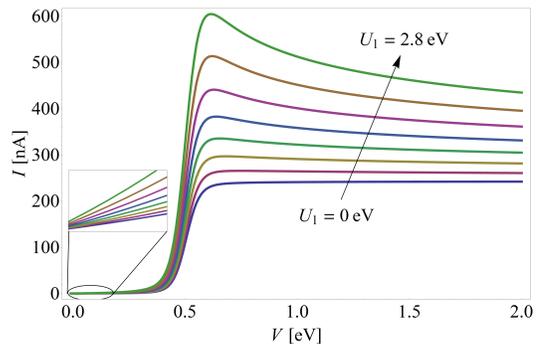}
\caption{I-V curves for the molecular junction for different values of the interface Coulomb interaction, $U_1=0,0.4,...,2.8$ eV (see text for other numerical parameters). As $U_1$ increases,  
the current develops a peak near the LUMO resonance and NDR. Inset: Blow-up of current at low bias, demonstrating an increase in the conductance with increasing $U_1$. }
\label{fig2}
\end{figure}

In Fig.~\ref{fig2}, the I-V curves are plotted for different values of the interface Coulomb term $U_1=0,0.4,0.8,...2.8$ eV, demonstrating explicitly the effect of 
the interface Coulomb interaction. With increasing interaction, the current develops a pronounced peak near the LUMO resonance and then decreases, 
resulting in NDR. In the inset of Fig.~\ref{fig2}, a blow-up of the current at low bias is plotted, showing that the interface interaction $U_1$ increases the 
current not only near the LUMO (resulting in NDR), but also in the linear response regime, due to the hybridization between the LUMO and the electrode states, even 
in the absence of current. For the parameters used here, this process results in a fourfold increase in the conductance. This implies that in molecular junctions 
with atomic-scale roughness, conductance may increase due to reduced screening of interactions in the electrodes (this seems to be in line with recent measurements \cite{NIJHUIS}).

\begin{figure}[h!]
\centering
\includegraphics[width=8truecm]{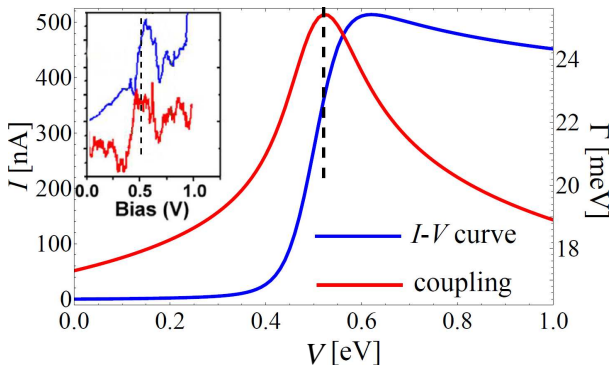}
\caption{Current (blue) and effective coupling $\Gamma$ (red) as a function of the bias voltage for the interface Coulomb interaction $U_1=2.6$eV. The coupling constant is maximal 
at the LUMO resonance, followed by a peek in the current. In the inset the experimental data of Ref.~\cite{Zhou2013} is shown, presenting the current (blue) and force constant (red). The dashed lines show the position of the maximum coupling, demonstrating the agreement between theory and experiment. }
\label{fig3}
\end{figure}

For a qualitative comparison with the experiment in Ref.~\cite{Zhou2013}, Fig.~\ref{fig3} shows both the I-V curve (blue solid line) and the effective coupling $\Gamma$ (red solid line) as a function of 
bias voltage, with $U_1=2.6$ eV. One can see that the effective $\Gamma$ is strongly bias dependent, with a maximal change where the bias reaches the LUMO energy (indicated by a dashed line). The 
inset shows the experimental results, where the I-V curve (blue) and the force constant (red), which is proportional to the coupling, are presented (adapted from Ref.~\cite{Zhou2013}). The dashed line 
in the inset indicates the maximum of the force, and the qualitative resemblance between the theoretical and the experimental curves is evident.

\begin{figure}[h!]
\centering
\includegraphics[width=9.5truecm]{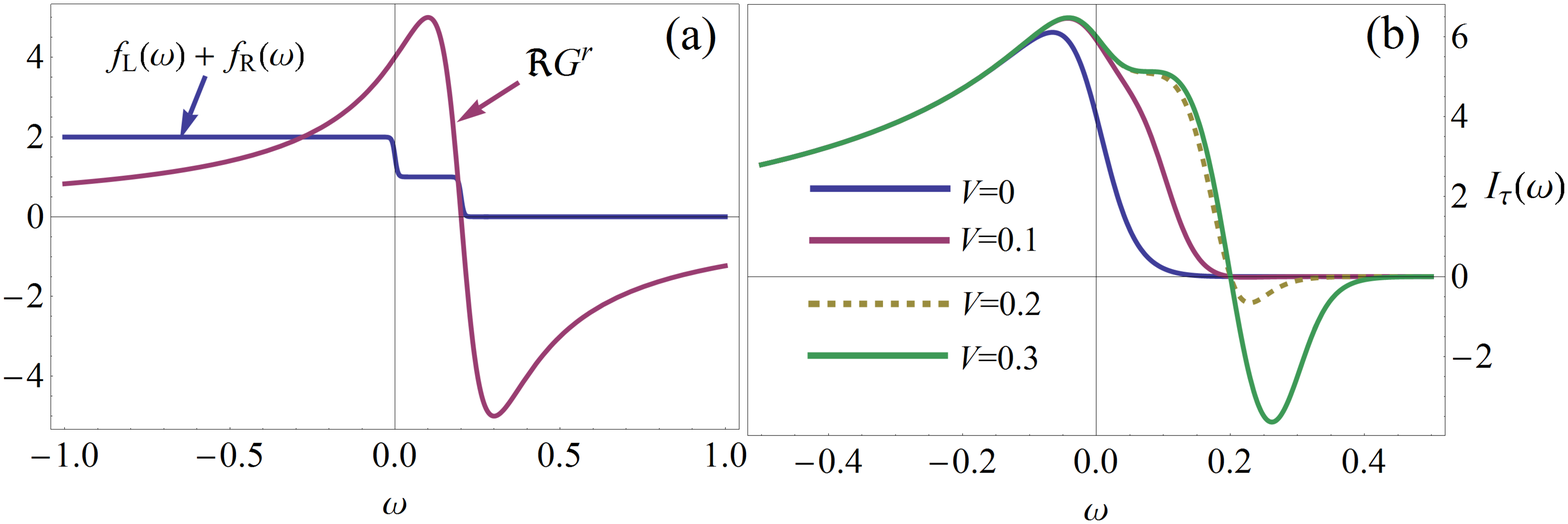}
\caption{(a) The parts $f_L(\omega)+f_R(\omega)$ (blue line) and $\Re G^r \omega$ (purple line) composing the integrand $I_\tau (\omega)$ (Eq.~\ref{Itau}), for a non-interacting example (see text). $\Re G^r \omega$ changes sign as the molecular level is crossed. (b)$I_\tau (\omega)$ for different bias voltages $V=0,0.1,0.2,0.3$. As long as $V<\e_0$, $I_\tau (\omega)$ is mostly positive, and reaches maximal area when $V=\e_0$ (dashed line), after which it develops a negative part, resulting in a decrease in $\tau$.}
\label{fig4}
\end{figure}

So it seems that one could explain the NDR in terms of an inhomogeneous dependence of the effective coupling on the bias voltage. but what is the origin of this inhomogeneity? 
To answer this, we look carefully at the expression for $\tau$ in Eq.~(\ref{subeq4}), and investigate its integrand,  
\beq I_\tau (\omega)= (f_L(\omega)+f_R(\omega)) \sum_\si\Re G^r_\si (\omega)~. 
\label{Itau}\eeq For the sake of simplicity we look at the integrand $I_\tau (\omega)$ for a non-interacting molecule, where the Green's function has the simple form $G^r=\frac{1}{\omega-\epsilon_0-i \Gamma}$, where $\epsilon_0$ is the molecular level and $\Gamma$ is the level broadening (spin-dependence was omitted). 
 In Fig.~\ref{fig4}(a) we plot the two parts of $I_\tau (\omega)$, i.e. $f_L(\omega)+f_R(\omega)$ and $\Re G^r \omega$ as a function of $\omega$ for $\e_0=0.2, ~\Gamma=0.1$ eV (at room temperature) and bias voltage 
$V=0.2$ eV, i.e. at the resonance. The most obvious feature is that $\Re G^r \omega$ changes sign as $\omega$ crosses $\e_0$. This can be understood as the fact that hole-like excitations ($\omega<\e_0$) having a 
positive contribution to the bound exciton amplitude, and electron-like excitations have a negative contribution to the amplitude of the bound exciton. In Fig.~\ref{fig4}(b) the integrand $I_\tau (\omega)$ is plotted for 
different bias voltages, $V=0,0.1,0.2,0.3$ eV. As long as $V<\epsilon_0$ only hole-like excitations are excited by $f_L(\omega)+f_R(\omega)$ and the integrand in positive, reaching a maximum when $V=\e_0$ (dashed line). Once $V$ 
crosses $\e_0$ there are negative contributions coming from the electron excitations, which reduce the integral, resulting in a decrease in $\tau$. 

To analyze in greater detail how the NDR depends on the structure of the molecular junction, we study the dependence of the peak-to-valley ratio (PVR), which is the ratio between the maximal current and 
the minimal current (but for increased bias, as implied by the NDR).  Specifically, we look at the PVR as a function of the distance $r$ between the molecule and the bridging site. This 
distance affects both the interface Coulomb interaction $U_1$ (via the Ohno parametrization) and the bare tunneling element $t$, which decreases exponentially with the distance 
as $t\approx t_0 e^{-\beta r}$ \cite{Wang2005} (here we take $\beta = 0.85$ \AA, but changing this parameter has only a quantitative effect on the results presented below). 
These two effects - the decrease in the interface Coulomb energy and the reduction of the tunneling element - compete with each other in determining the magnitude of the NDR, since the 
smaller the bare tunneling matrix element, the larger the relative effect of the interface Coulomb interaction.

\begin{figure}[h!]
\centering
\includegraphics[width=8truecm]{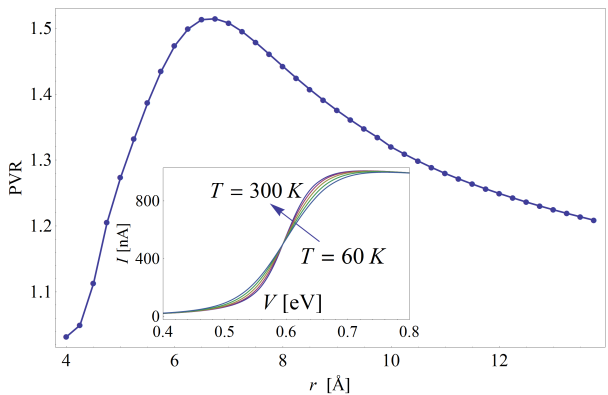}
\caption{Peak-to-valley ratio (PVR) as a function of the molecule-electrode distance $r$, which affects both the interface interaction $U_1$ and the bare tunneling elements $t$ (see text). The strong dependence of the PVR on $r$ can be tested experimentally, and is in accord with the experimental findings. Inset: I-V curves for different temperatures $T=60,120,...,300$ K, demonstrating a very weak dependence of the current on the temperature. This weak dependence can also serve as a test for the origin of the NDR. }
\label{fig5}
\end{figure}

In Fig.~\ref{fig5} the PVR vs. $r$ is plotted, and a well-defined maximum is observed, demonstrating 
this competition. Note that for short distances the PVR tends to unity, implying that the NDR vanishes, and the same is true for large $r$. This relationship may explain the experimental observation \cite{Zhou2013} that only 
some junctions exhibit NDR, which is correlated with their conductance (and hence with their molecule-interface geometry and the values of the bare coupling and the interface Coulomb 
interaction). Fig.~\ref{fig5} predicts that within a specific molecular junction, the PVR can be reduced or increased by squeezing or stretching the junction respectively, which is well within  current experimental reach \cite{Zhou2011,Zhou2013}.   

Another experimental test which may help differentiating between possible origins for the NDR is the temperature dependence of the current (and resulting PVR). The results of Ref.~\cite{Zhou2013} show a molecular junction with PVR of $\sim 1.5$, similar to that of Fig.~\ref{fig5}. In Ref.~
\cite{Chen2000} it was shown that molecular junctions that exhibit $PVR\sim 1.5$ at room temperature can have either similar or very high ($\sim 1000)$) PVR at low 
temperatures, depending on the structure of the 
molecular junction. Thus, the different mechanisms that give rise to NDR may be differentiated from one 
another by their temperature dependence. Within the calculation presented here, temperature appear explicitly only via the Fermi functions, yet due to the coupling renormalization, the level broadening $\Gamma$ also depends on temperature. In the inset of Fig.~\ref{fig5}, we plot the I-V curves for the 
same parameters as in Fig. ~\ref{fig3} for temperatures $T=60,180,...,300$ K. As may be seen, the NDR 
depends very weakly on temperature. This suggests that in situations in which the NDR is determined by a 
change in 
the coupling (as opposed to a shift in the transmission resonances) there will not be a substantial difference in the PVR between low and high temperatures. 
 
\begin{figure}[h!]
\centering
\includegraphics[width=7truecm]{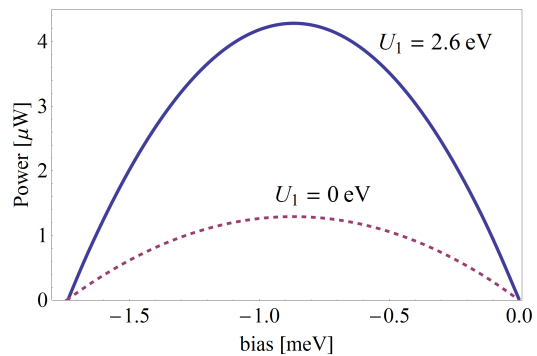}
\caption{Power output vs. bias voltage of the molecular junction in the presence of a constant temperature gradient $\Delta T=50$ K, for $U_1=2.6$ eV (blue solid line) and 
$U_1=0$ eV (dashed red line). }
\label{fig6}
\end{figure}

The coupling between the molecular levels and the electrode levels is essential not only for the I-V characteristics of molecular junctions but also for the thermal transport and the thermo-electricity in molecular junctions, which have been central research themes in recent years \cite
{Mahan1996,Paulsson2003,Segal2005,Nitzan2007,Malen2010,Dubi2011,Tsutsui2012}. The dynamical coupling renormalization mechanism, described here to address the NDR in molecular junctions, may play a role in 
determining, for example, the thermoelectric conversion properties of such junctions. To demonstrate this, we study the response of the molecular junction to a constant temperature 
difference $\Delta T=50$ K (obtained by keeping $T_L$ at room temperature and increasing $T_R$ by $\Delta T$). When a temperature difference is present, a current $I_{sc}$ 
starts flowing through the junction. Applying a bias voltage gradually reduces the current until at a voltage $V_{oc}$, the 
current vanishes. At both vanishing current or zero bias, the power output of the junction vanishes, and in between the power output reaches a maximum, and that point is considered the preferable 
working point for thermoelectric junctions \cite{Esposito2009}. Since the junction is strongly out of equilibrium at maximum power (as opposed to the linear response 
regime), it is expected that there will be a large renormalization of the coupling, which will affect the maximal power. In Fig.~\ref{fig6} the power  output $P=I \times V$ 
of the junction with the parameters of Fig. ~\ref{fig3} at a constant temperature difference of $\Delta T=50$ K is plotted as a function of the voltage bias, for two values 
of the interface Coulomb interaction $U=0$ eV (dashed line) and $U=2.6$ eV (solid line). We find that while $V_{oc}$ is unchanged (as expected), the maximal power depends strongly on the 
interface Coulomb interaction, which increases it by a factor of three. 

\section{Summary and Conclusion} 
To summarize, we suggested that the Coulomb interaction across the molecule-electrode interface induces a 
dynamical, bias-dependent renormalization of the molecule-electrode coupling. The origin of the change in the coupling is the binding of electrons in the molecule and holes 
in the electrode atoms closest to the molecule. The effective coupling has a maximum when the bias reaches the LUMO energy level, resulting in 
an inhomogeneous I-V curve and a negative differential resistance. The theoretical features, including the relation between the coupling inhomogeneity and the NDR and changes in 
the PVR are in accord with recent experimental findings. We suggested various experiments to test the origin of the NDR and predicted their outcome. Finally, 
we demonstrated that the mechanism we suggest affects other transport properties, such as thermoelectric conversion power output. 
 
We conclude by noting that most current theories of transport through molecular junctions use a method combining density functional theory (DFT) and non-equilibrium Green's 
functions, where the electrodes are treated as non-interacting far from the molecule itself. The self-consistent correction to the coupling that was introduce here, (\ref{subeq1}-\ref{subeq4}), can in principle be inserted into these calculations with parameters calculated within the DFT scheme. This provides a correction
to transport calculations of molecular junctions that may prove important in evaluating transport properties from first principles.  
\begin{acknowledgments}
This research was funded by a Ben-Gurion University start-up grant.
\end{acknowledgments}

\appendix
\section{Derivation of the equation for $\tau$}
 In this appendix we provide the details of the calculation leading to the self-consistent equations Eq.~(3) in the main text. The starting point is the typical Hamiltonian for a molecular junction \cite{Tsutsui2012,Selzer2006,DiVentra2008}
\beqa
\cH &=& \cH_L+\cH_R+\cH_M+\cH_T+\cH_{I} \nonum
\cH_{L,R} &=& \sum_{k\si} \e_k c^{\dagger}_{k\si,L/R} c_{k\si,L/R} \nonum 
\cH_M &=& \e_0  \sum_\si d^{\dagger}_\si d_\si + U_0 d^{\dagger}_\uparrow d_\uparrow d^{\dagger}_\downarrow d_\downarrow  \nonum 
\cH_T &=& -\tilde{t}\sum_\si \left( c^{\dagger}_{\si,L} d_\si+ c^{\dagger}_{\si,R} d_\si+\textrm{h.c.} \right) \nonum
\label{Hamiltonian2} \eeqa  
where $c^{\dagger}_{k,L/R}$ creates an electron with momentum $k$ and spin $\si$ at the left/right electrode, $d^\dagger_\si$ creates an electron with spin $\si$ 
at the LUMO (and $c_{k,L/R}$ and $d_\si$ annihilate the same). The first three terms are a typical description of transport through molecular junctions, with metallic electrodes, molecular orbital 
energy $\e_0$, and tunneling matrix element $-t$ connecting the molecule and the electrodes. This Hamiltonian is already the mean-field version of the interaction Hamiltonian, where the interaction was 
decoupled into the the effective tunneling $\tilde{t}=t+\tau$, where $\tau$ is defined as the correlation function $\tau_{L,d}=U_1 \langle c^\dagger_{L,\si} d_\si \rangle,~ \tau_{d,L}=U_1 \langle  
d^\dagger_\si c_{L,\si} \rangle$ (we will show that the subscripts can be discarded). 

To proceed we write the correlation functions in term of the equal-time Keldysh lesser Green's functions
\beqa
 \langle c^\dagger_{L,\si} d_\si \rangle &=& \frac{-i}{2 \pi}\int d\omega G^<_{Ld,\si}(\omega)\nonum \langle  d^\dagger_\si c_{L,\si} \rangle &=& \frac{-i}{2 \pi}\int d\omega G^<_{dL,\si}(\omega)~~.
\label{eq2}\eeqa 

For these Green's functions, since the electrodes are non-interacting, one can write the Dyson's equations (we omit the spin subscript and explicit $\omega$ dependence)
\beqa 
G^<_{Ld}&=& -t\sum_k (g^<_k G^t-g^{\bar{t}}_k G^<) \nonum 
G^<_{Ld}&=& -t\sum_k (g^t_k G^<-g^<_k G^{\bar{t}}) 
\label{eq3}\eeqa
Where $g$ and $G$ are the Green's function for the electrodes and the molecular electrons respectively, and $t$ and $\bar{t}$ are the time-ordered and anti-time-ordered Keldysh functions. 
Since $G^<_{Ld}=-(G^<_{dL})^*$ and noting the analytic properties of the keldysh functions (which guarantees that the real part will vanish), one can study $\tau=\frac{1}{2}(\tau_{L,d}+\tau_{d,L})$. 
Thus we use 
\beq G^<_{Ld}+G^<_{dL}=-t\sum_k \left( g^<_k (G^t-G^{\bar{t}})+G^<(g^t-g^{\bar{t}}) \right) ~~. \eeq The Keldysh functions obey the relation $G^t-G^{\bar{t}}=G^r+G^a$, and similarly for $g$. 
For the electrodes we use the wide-band approximation, which sums up to the relation $\sum_k g^{r,a}_k=\pm i \Gamma$, and so 
\beq 
G^<_{Ld}+G^<_{dL}=-t\sum_k  g^<_k (G^t-G^{\bar{t}})~~.
\eeq
The lesser function for the leads obeys $\sum_k g^<_k=-2\pi i \rho_0 f_L(\omega)$, which gives finally 
\beq 
G^<_{Ld}+G^<_{dL}=-4\pi i t \rho_0 f_L(\omega) \Re G^r~~.
\eeq
Going back to the definition of the correlation functions we have 
\beq 
\frac{1}{2} (\langle c^\dagger_{L,\si} d_\si \rangle+\langle  d^\dagger_\si c_{L,\si} \rangle) =-i\times (-2 \pi i t \rho_0 \int \frac{d\omega}{2 \pi} f_L (\omega ) \Re G^r(\omega) 
\eeq 
which gives the final equation for $\tau$ (averaging over left and right electrodes) ,  
\beq 
\tau=-\frac{1}{2} U_1 t \rho_0 \int d\omega (f_L(\omega )+f_R (\omega  \Re G^r (\omega )~~.
\eeq


\end{document}